\documentclass{emulateapj}
\usepackage{apjfonts}

\newcommand{\xivec}{\mbox{\boldmath $\xi$}}

\lefthead{HAN} 
\righthead{MICROLENSING ZONE}

\begin{document}
\title{Microlensing Zone of Planets Detectable through the 
Channel of High-Magnification Events}

\author{Cheongho Han}
\affil{Program of Brain Korea 21, Department of Physics, 
Chungbuk National University, Chongju 361-763, Korea;\\
cheongho@astroph.chungbuk.ac.kr}

\submitted{Submitted to The Astrophysical Journal}

\begin{abstract}
A microlensing lensing zone refers to the range of planet-star 
separations where the probability of detecting a planetary signal 
is high. Its conventional definition as the range between $\sim 0.6$ 
and 1.6 Einstein radii of the primary lens is based on the criterion 
that a major caustic induced by a planet should be located within 
the Einstein ring of the primary.  However, current planetary lensing 
searches focus on high-magnification events to detect perturbations 
induced by another caustic located always within the Einstein ring 
very close to the primary lens (stellar caustic) and thus a new 
definition of a lensing zone is needed. In this paper, we determine 
this lensing zone.  By applying a criterion that detectable planets 
should produce signals $\geq 5\%$, we find that the new lensing zone 
varies depending on the planet/star mass ratio unlike the fixed range 
of the classical zone regardless of the mass ratio.  The lensing zone 
is equivalent to the classical zone for a planet with a planet/star 
mass ratio $q\sim 3\times 10^{-4}$ and it becomes wider for heavier 
planets.  For a Jupiter-mass planet, the lensing zone ranges from 
0.25 to 3.9 Einstein radii, corresponding to a physical range between 
$\sim 0.5$ AU and 7.4 AU for a typical Galactic event. The wider 
lensing zone of central perturbations for giant planets implies that 
the microlensing method provides an important tool to detect planetary 
systems composed of multiple ice-giant planets.
\end{abstract}



\section{Introduction}

The microlensing signal of a planet is a perturbation to the smooth 
standard light curve of the primary-induced lensing event occurring 
on a background source star.  The planetary signal lasts for a short 
period of time: several days for a gas giant planet and several hours 
for an Earth-mass planet.  Currently, the observational frequency of 
lensing surveys is $\sim 1$ per night and thus it is not enough to 
detect planetary signals.  To achieve the observational frequency 
required to detect planetary signals, microlensing planet searches 
are using a combination of survey and follow-up observations.  Survey 
observations (e.g., OGLE, {http://www.astrouw.edu.pl/$\sim$ogle}, 
\citet{udalski03}; MOA, http://www.physics.canterbury.ac.nz/moa/,
\citet{bond02a}) aim to maximize the event rate by monitoring a 
large area of the Galactic bulge field on a roughly nightly basis.  
They issue alerts of ongoing events in the early stage of lensing 
magnification by analyzing data in real time.  Follow-up observations 
(e.g., PLANET, http://planet.iap.fr, \citet{kubas08}; MicroFUN, 
http://www.astronomy.ohio-state.edu/\%7Emicrofun/, \citet{dong06}) 
are focused on the alerted events in order to detect short-lived 
planetary signals.

\begin{figure*}[t]
\epsscale{0.85}
\plotone{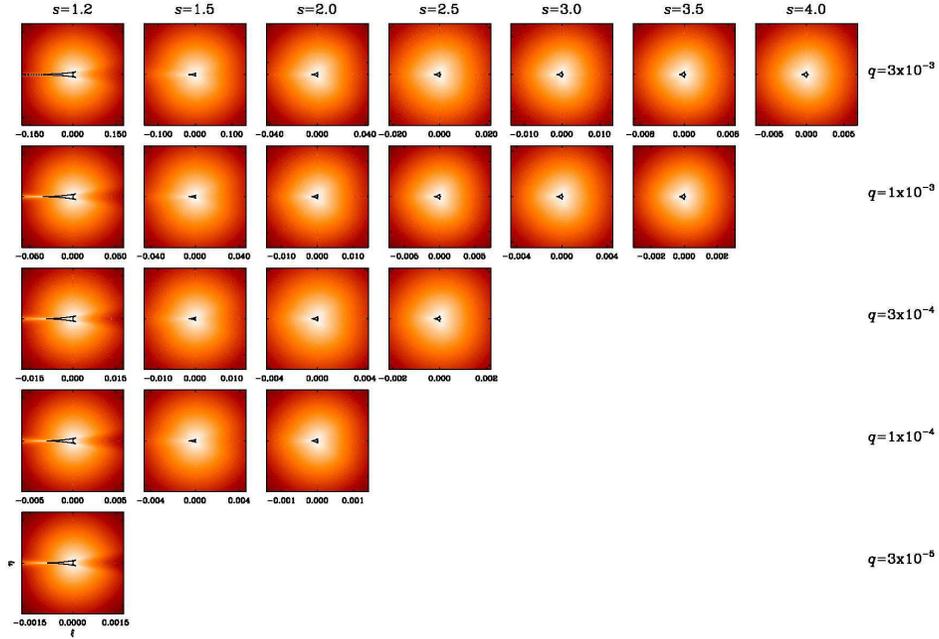}
\caption{\label{fig:one}
Maps of magnification pattern involved with a point source in the 
central region of planetary systems with various planet-star 
separations $s$ and planet/star mass ratio $q$.  The coordinates 
$(\xi,\eta)$ represent the axes that are parallel with and normal 
to the planet-star axis, respectively.  In each map, the star is 
at the center and the planet is on the left.  The closed figures 
drawn in black curves represent the caustics.  Greyscale is drawn 
such that brighter tone represents the region of a higher 
magnification.   
}\end{figure*}

In practice, however, the number of telescopes available for 
follow-up observations is far less for intensive followups of all 
alerted events and thus priority is given to those events which will 
maximize the planetary detection probability.  Currently, the highest 
priority is given to high-magnification events \citep{bond02b, abe04, 
rattenbury02}.  This is based on the fact that in addition to a
major caustic located away from the primary lens (planetary caustic), 
a planet induces an additional tiny caustic in the region close to 
the primary lens (stellar or central caustic).  Due to its location, 
the perturbations induced by the stellar caustic always occur near 
the peak of high-magnification events and this makes follow-up 
observations right on the period of greatest sensitivity possible.  
The strategy focusing on high-magnification events was proposed by 
\citet{griest98} and the adoption of the strategy has led to the 
discoveries of six planet candidates (OGLE-2005-BLG-071Lb, 
OGLE-2005-BLG-169Lb, OGLE-2006-BLG-109Lb,c, MOA-2007-BLG-400b, 
MOA-2007-BLG-192) among the total eight microlensing planet 
candidates discovered to date \citep{bond04, udalski05, beaulieu06, 
gould06, gaudi08, dong08, bennett08}.

In planetary microlensing, a `lensing zone' refers to the range of 
planet-star separations where the probability of detecting a
planetary signal is high.  It was first discussed by \citet{gould92},
who found that the planet detection probability is maximized when 
the planet is near the Einstein radius of the primary star.  Later, 
the definition of the lensing zone is refined by \citet{wambsganss97} 
and \citet{griest98} as the range between approximately 0.6 and 1.6 
Einstein radii of the primary lens.  In defining the lensing zone, 
the location of a caustic is important because perturbations can be 
covered by follow-up observations when they occur during the progress 
of the lensing magnification induced by the primary star.\footnote{We
note that events are occasionally monitored in current follow-up 
observations even after the source has exited the Einstein ring.
An example is the event OGLE-2005-BLG-390 \citep{beaulieu06} for
which the planet was detected at the moment when the source was 
about to exit the ring. For these events, the lensing zone is more 
extended and broader than the classical definition of the lensing zone.} 
When a planet is located within this range, not only the size of the 
planet-induced caustic is maximized but also the caustic is located 
inside the Einstein ring of the primary lens.  However, this definition 
of the lensing zone is based only on the location of the planetary 
caustic.  Considering that current lensing observations focus on 
perturbations induced by stellar caustics and these caustics are 
always located within the Einstein ring regardless of the planet-star 
separation, a new definition of the lensing zone is needed.  In this 
paper, we determine the range of a lensing zone for microlensing 
planets detectable through the channel of high-magnification events.

The paper is organized as follows.  In \S\ 2, we briefly describe 
the properties of planet-induced caustics.  In \S\ 3, we investigate 
the patterns of lensing magnifications in the central region of 
planetary lens systems with various planet/star mass ratios and 
planet-star separations.  From the investigation of the magnification 
pattern, we set the criterion for planet detections and determine 
the lensing zone based on this criterion. We discuss about the 
implication of the determined lensing zone.  We summarize the 
result and conclude in \S\ 4.

\section{Caustics of Planetary Lensing}

A planetary lensing corresponds to the case of binary lensing with a 
very low mass companion.  One important characteristic of binary 
lensing is the formation of caustics.  Caustics represent the set of 
source positions at which the magnification of a point source becomes 
infinite.  For a planetary case, there exist two sets of disconnected 
caustics. One small stellar caustic is located near the primary star 
and the other planetary caustic is located away from the star.

The position vector to the center of the planetary caustic from the
position of the primary lens is  
\begin{equation}
\xivec_{pc} = {\bf s}\left( 1-{1\over s^2}\right),
\label{eq1}
\end{equation}
where ${\bf s}$ is the position vector of the planet from the star 
normalized by the Einstein radius of the primary, $r_{\rm E}$.  
The Einstein radius is related to the physical parameters of the 
lens system by
\begin{equation}
r_{\rm E}\sim 3.8\ {\rm AU}\ 
\left( {M\over 0.3\ M_\odot} \right)^{1/2}
\left( {D_L\over 6\ {\rm kpc}} \right)^{1/2}
\left(  1-{D_L\over D_S}\right)^{1/2},
\label{eq2}
\end{equation}
where $M$ is the mass of the primary lens and $D_L$ and $D_S$ are 
the distances to the lens and source star, respectively.  Then, 
the planetary caustic is located within the Einstein ring for 
planets located in the range of 
\begin{equation}
{\sqrt{5}-1 \over 2} \leq s \leq {\sqrt{5}+1 \over 2}.
\label{eq3}
\end{equation}
This range corresponds to the lensing zone defined by 
\citet{wambsganss97} and \citet{griest98}.  
More details about the characteristics 
of planetary caustics are found in \citet{han06}.

The stellar caustic is smaller than the planetary caustic.
Its size as measured by the width along the planet-star axis
is related to the planet/star mass ratio $q$ and planet-star 
separation $s$ by
\begin{equation}
\Delta\xi_{cc} \sim {4q\over (s-s^{-1})^2}.
\label{eq4}
\end{equation}
The size of the stellar caustic decreases as $\propto q$, while the 
size of the planetary caustic decreases as $\propto q^{1/2}$.  This 
implies that the stellar caustic shrinks more rapidly than the 
planetary caustic as the mass ratio decreases.  In the limiting case 
of a very wide planet ($s\gg 1$) and a close-in planet ($s\ll 1$), 
the dependency of the size of the stellar caustic on the planet 
separation is 
\begin{equation}
\Delta \xi_{cc} \propto 
\cases{
s^{-2} & for $s\gg 1$, \cr
s^{2} & for $s\ll 1$. \cr
}
\label{eq5}
\end{equation}
For a given mass ratio, a pair 
of stellar caustics with separations $s$ and $s^{-1}$ are identical 
to the first order of perturbation approximation.  Although the size 
depends both on $q$ and $s$, the shape of the stellar caustic is 
solely dependent on the planet separation.  
For more detailed characteristics of stellar caustics, 
see \citet{chung05}.

\begin{figure*}[t]
\epsscale{0.85}
\plotone{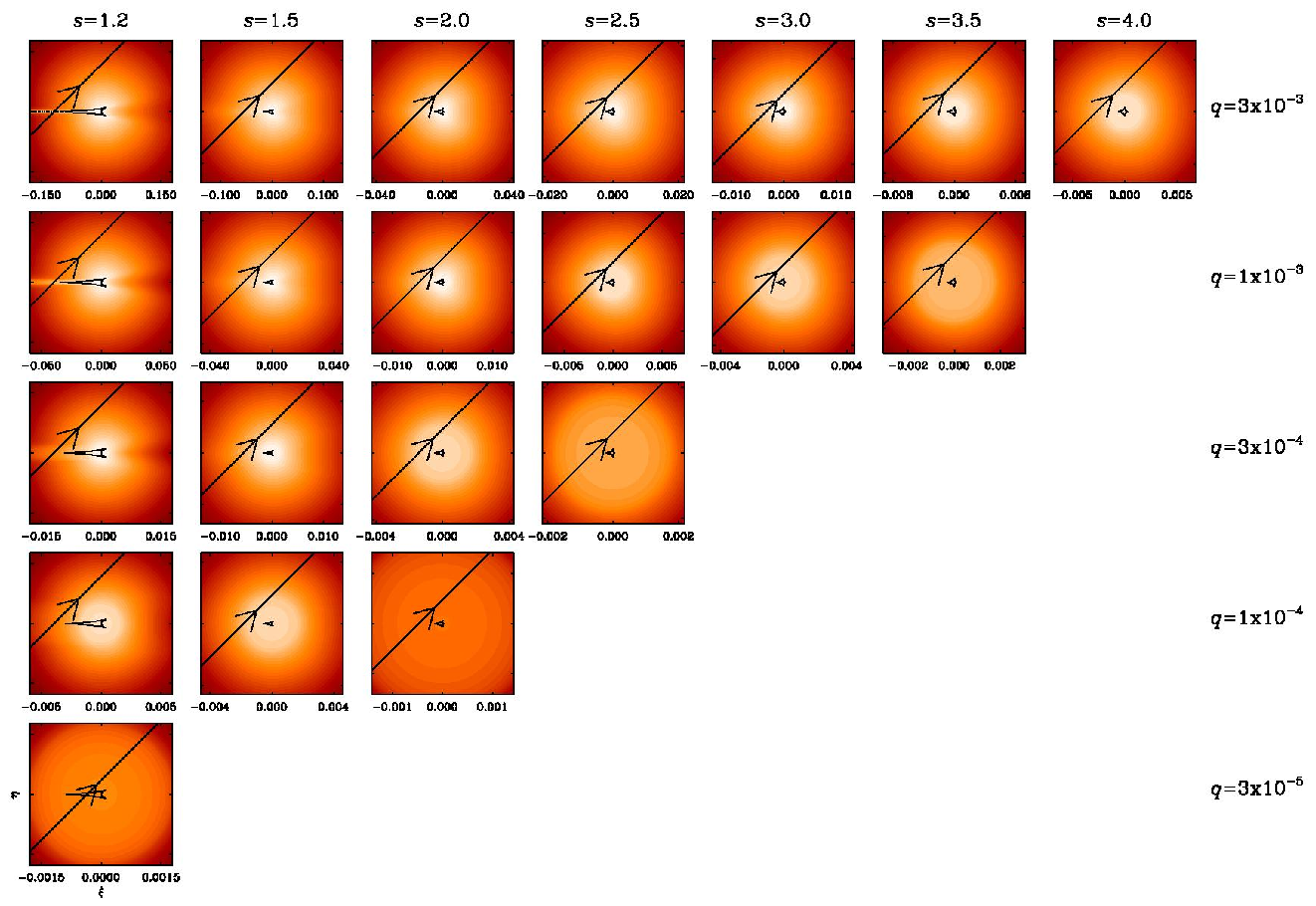}
\caption{\label{fig:two}
Magnification patterns with finite-source effect.  Notations are same 
as in Fig.~\ref{fig:one}.  The straight lines with arrows represent 
the source trajectories and the light curves of the resulting events 
are presented in the corresponding panels of Fig.~\ref{fig:three}.  
}\end{figure*}

\begin{figure*}[t]
\epsscale{0.85}
\plotone{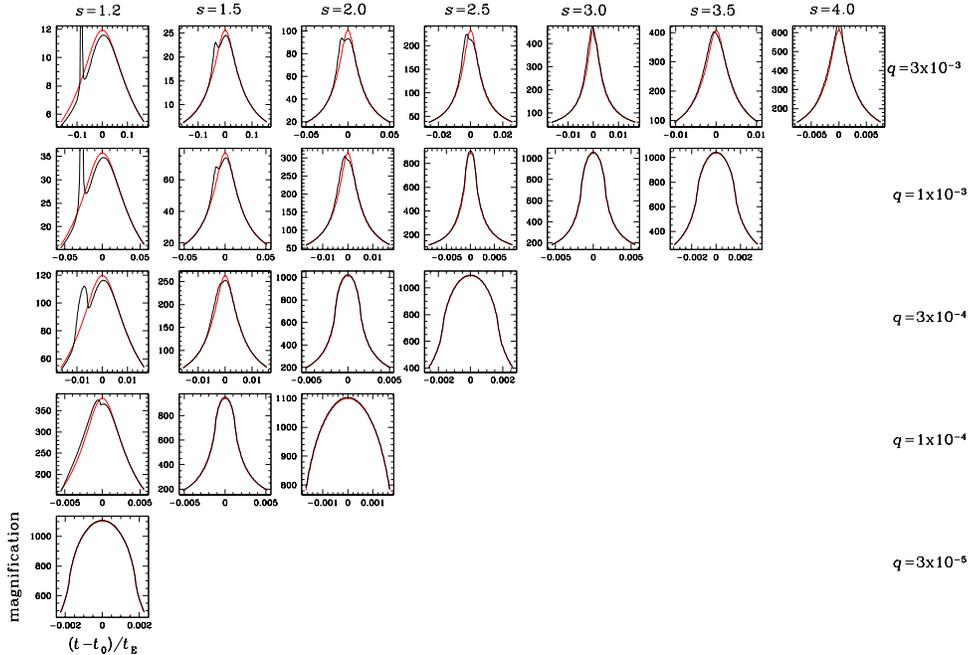}
\caption{\label{fig:three}
Example light curves of planetary lensing events with high 
magnifications.  The lens systems and trajectories responsible for 
the individual light curves are presented in the corresponding panels 
in Fig.~\ref{fig:two}.  Black and red curves are for events with 
and without the planet, respectively.
}\end{figure*}

\section{Central Perturbations}

\subsection{Magnification Pattern}

Since the stellar caustic is always close to the primary lens, its 
location can no longer be a criterion for the the determination 
of the lensing zone for planets detectable through the channel of 
high-magnification events.  For these planets, the most important 
restriction on planet detection is the size of the caustic.  If the 
caustic is too small, the perturbation produced by the caustic is 
severely washed out by the finite-source effect and thus cannot be 
detected \citep{bennett96}.  Since the stellar caustic is small, 
perturbations induced by the stellar caustic are especially vulnerable 
to the finite-source effect.

\begin{figure*}[t]
\epsscale{0.85}
\plotone{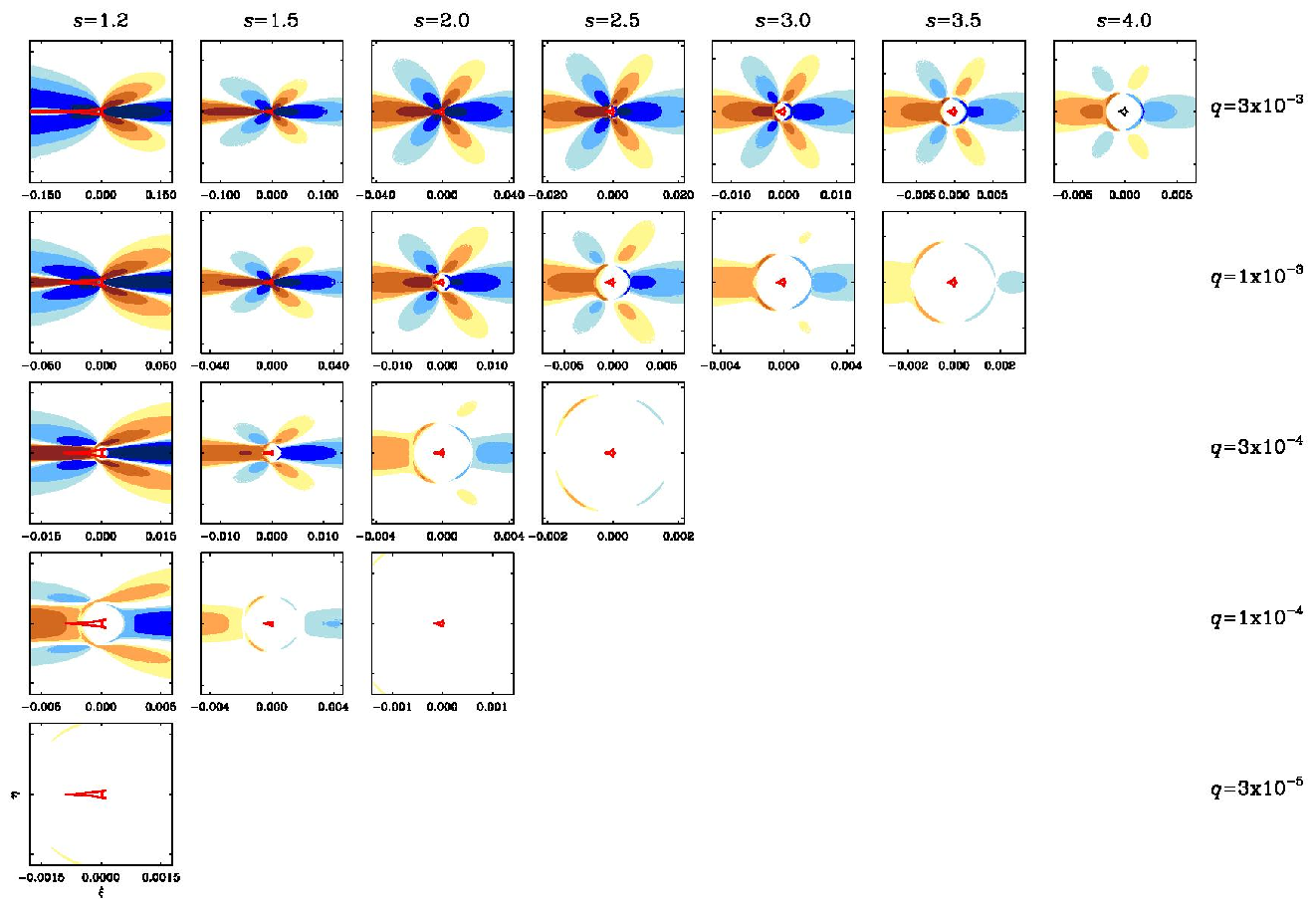}
\caption{\label{fig:four}
Maps of magnification excess $\epsilon$.  In each map, the regions 
with bluish and brownish colors represent the areas where the 
planetary lensing magnification is lower and higher than the 
single-lensing magnification, respectively.  The individual 
colored areas represent the regions with $|\epsilon|\geq 2\%$, 
3\%, 5\%, and 10\%, respectively, with thickening tones of color.   
}\end{figure*}

To see how the finite-source effect affects planetary perturbations,
we construct maps of magnification pattern in the central region of  
planetary systems with various mass ratios and separations.  The 
constructed maps are presented in Figure~\ref{fig:one} and Figure 
\ref{fig:two}.  For comparison, we present two sets of maps where 
the maps presented in Figure~\ref{fig:one} are constructed not 
considering the finite-source effect, while those presented in 
Figure~\ref{fig:two} are constructed by considering the effect.  
The finite-source effect is parameterized by the ratio of the 
source radius, $r_\star$, to the Einstein radius of the primary 
lens.  For Galactic bulge events, this ratio is scaled by the 
physical parameters of the lens system as
\begin{equation}
\rho_\star = 9\times 10^{-4} 
\left( {r_\star\over R_\odot}\right)
\left[
\left( {M\over 0.3\ M_\odot}\right)
\left( {D_L\over 6\ {\rm kpc}}\right)
\left( 1-{D_L\over D_S}\right)\right]^{-1/2}.
\label{eq6}
\end{equation}
We consider the finite-source effect by assuming that the source 
star has a uniform disk with  a radius equivalent to the sun and 
the physical parameters of the lens system are $M=0.3\ M_\odot$, 
$D_L=6$ kpc, and $D_S=8$ kpc by adopting the values of a typical 
Galactic bulge event.  This results in $\rho_\star =1.8\times 10^{-3}$.  
Then, the magnification affected by the finite-source effect becomes
\begin{equation}
A={
\int_0^{\rho_\star} I(r)A_p(|{\bf r}-{\bf r}_L|) r dr
\over
\int_0^{\rho_\star} I(r) r dr
},
\label{eq7}
\end{equation}
where ${\bf r}_L$ is the displacement vector of the source center 
with respect to the lens, ${\bf r}$ is the vector to a position 
on the source star surface with respect to the center of the source 
star, $I(r)$ represents the surface brightness profile of the source 
star, and $A_p$ is the point-source magnification.  We note that the 
blank space with missing panels in Figure~\ref{fig:one} and 
Figure~\ref{fig:two} implies that the magnification pattern is difficult 
to be distinguished from that of a single-lensing case.  We also note 
that maps are presented only for the cases with $s>1.0$ because the 
magnification patterns for a pair of planetary systems with $s$ and 
$s^{-1}$ are identical \citep{dominik99}.

In Figure~\ref{fig:three}, we present example light curves of 
high-magnification events.  The source trajectories responsible for 
the individual events are marked in the corresponding panels of 
Figure~\ref{fig:two}.  In each panel, there are two light curves.  
One is for planetary lensing event (black curve) and the other is 
for a single-lensing event without the planet (red curve).

Also presented in Figure~\ref{fig:four} are the maps of magnification 
excess, which represents the fractional deviation of the planetary 
lensing magnification $A$ from the single-lensing magnification $A_0$, 
i.e.
\begin{equation}
\epsilon = {A-A_0  \over A_0}.
\label{eq8}
\end{equation} 
We note that the finite-source effect is taken into consideration 
for both $A$ and $A_0$.  In each excess map, the region with bluish 
colors represents the area where the planetary lensing magnification 
is lower than that of the single lensing, i.e.\ $\epsilon < 0$,  
while the region with brownish colors represents the area where the 
planetary lensing magnification is higher, i.e.\ $\epsilon > 0$.  
The individual colored areas represent the regions with  $|\epsilon| 
\geq 2\%$, 3\%, 5\%, and 10\%, respectively, with thickening tones 
of color.

\subsection{Lensing Zone}

The pattern of central perturbations depends both on the shape of
the stellar caustic and its relative size to that of the source 
star.  We conduct detailed investigation about the dependencies 
of perturbation pattern on the caustic shape and size.  The 
results of this investigation are themselves very important for the 
understanding of central perturbations and thus we are preparing 
a separate paper for that.  Although the detailed results of this 
work are not presented in this paper, several findings relevant 
to this work are (1) the dependency of the finite-source effect 
on the caustic shape is weak and (2) perturbations persist even 
when the caustic is substantially smaller than the source size.  
Specifically, we find that perturbations with $\epsilon\geq 5\%$ 
can be detected if the caustic is bigger than roughly one fourth 
of the source size (diameter), i.e., $\Delta\xi_{cc} /2\rho_\star
\gtrsim 1/4$.  This result is consistent with the case of a 
recently reported microlensing planet MOA-2007-BLG-400b that was 
discovered through the high-magnification channel \citep{dong08}.  
In this case, the caustic width is $\Delta \xi_{cc}=0.0015$ and the 
source diameter is $2\rho_\star = 0.0064$ and thus the ratio of the 
caustic width to the source diameter is $\Delta\xi_{cc}/2\rho_\star 
\sim 0.23$.  By adopting the criterion of planet detection as 
$\Delta\xi_{cc}/2\rho_\star\gtrsim 1/4$ and with the expression of 
the caustic size in equation~(\ref{eq4}), the condition for planet 
detection is 
\begin{equation}
{8q\over \rho_\star (s-s^{-1})^2}\gtrsim 1,
\label{eq9}
\end{equation}
or equivalently, 
\begin{equation}
s^2-\sqrt{8q\over \rho_\star}s-1\lesssim 0.
\label{eq10}
\end{equation}
Solving this relation results in an analytic expression for the 
lensing zone of central perturbations of 
\begin{equation}
\left\vert
\sqrt{{2q\over \rho_\star}} - \sqrt{{2q\over \rho_\star}+1}
\right\vert
\lesssim s \lesssim
\sqrt{{2q\over \rho_\star}} + \sqrt{{2q\over \rho_\star}+1}.
\label{eq11}
\end{equation}

\begin{figure}[t]
\epsscale{1.2}
\plotone{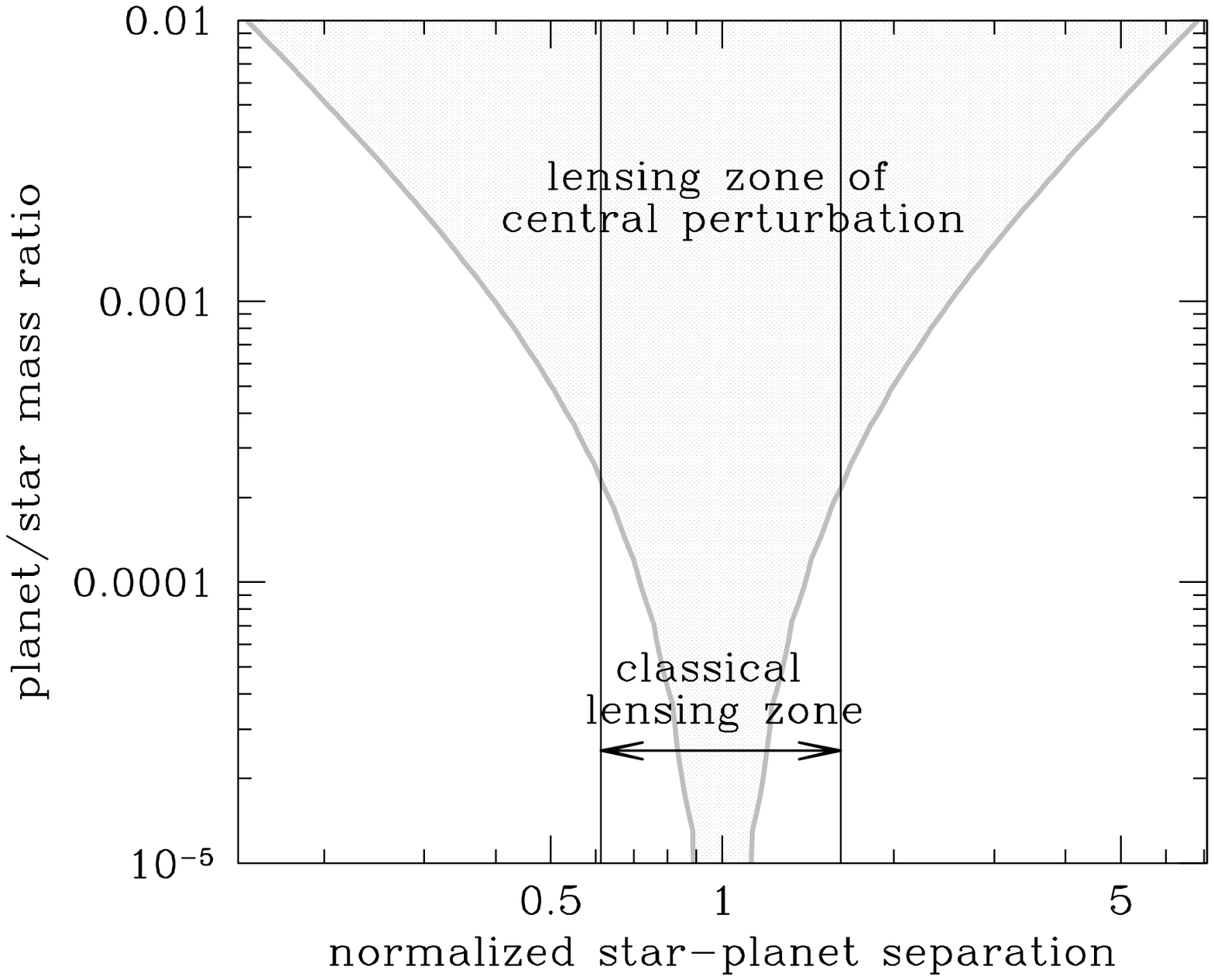}
\caption{\label{fig:five}
Lensing zone (shaded area) of planets detectable through the channel 
of high-magnification events in the parameter space of $(s,q)$. 
}\end{figure}

\begin{deluxetable}{ccc}
\tablecaption{Lensing Zones\label{table:one}}
\tablewidth{0pt}
\tablehead{
\multicolumn{1}{c}{planet/primary} &
\multicolumn{2}{c}{lensing zone} \\
\multicolumn{1}{c}{mass ratio} &
\multicolumn{1}{c}{in normalized units} &
\multicolumn{1}{c}{in physical units} 
}
\startdata
$3\times 10^{-3}$ & $0.26 \lesssim s \lesssim 3.9$ &  $0.5\ {\rm AU} \lesssim d \lesssim 7.4\ {\rm AU}$ \\
$1\times 10^{-3}$ & $0.40 \lesssim s \lesssim 2.5$ &  $0.8\ {\rm AU} \lesssim d \lesssim 4.8\ {\rm AU}$ \\
$3\times 10^{-4}$ & $0.58 \lesssim s \lesssim 1.7$ &  $1.1\ {\rm AU} \lesssim d \lesssim 3.3\ {\rm AU}$ \\
$1\times 10^{-4}$ & $0.72 \lesssim s \lesssim 1.4$ &  $1.4\ {\rm AU} \lesssim d \lesssim 2.6\ {\rm AU}$ \\
$3\times 10^{-5}$ & $0.83 \lesssim s \lesssim 1.2$ &  $1.6\ {\rm AU} \lesssim d \lesssim 2.3\ {\rm AU}$  
\enddata 
\tablecomments{ 
Lensing zones of central perturbations for several example planets. 
The physical ranges are estimated by assuming the Einstein radius is 
$r_{\rm E}=1.9$
AU. 
}
\end{deluxetable}

In Figure~\ref{fig:five}, we present the determined lensing 
zone (shaded area) in the parameter space of $(s,q)$.  From the 
comparison of the lensing zone of central perturbations with that 
of perturbations produced by planetary caustics, it is found that 
the lensing zone of central perturbations varies depending on the 
mass ratio unlike the fixed range of the classical lensing zone 
regardless of the mass ratio.\footnote{Strictly speaking, the 
lensing zone of perturbations induced by planetary caustics also
depends 
on the mass ratio.  This is because the planetary caustic has a 
finite size and thus part of the perturbation region induced by 
the caustic can be located within the Einstein ring even when the 
caustic center is outside the Einstein ring.  The caustic size and 
the resulting perturbation region increase with the increase of 
the mass ratio and thus the lensing zone becomes broader as the 
mass ratio increases.  However, we note that the caustic is small 
and thus this dependency is weak.  For a planet with a mass ratio 
$q=10^{-3}$, we find that this effect makes the lensing zone 
increase by $\sim 7\%$ assuming that the perturbation region 
extends 4 times of the caustic size.}  This is because while the 
position of the planetary caustic, which is the criterion for 
the classical lensing zone, does not depend on the mass ratio, 
the size of the stellar caustic, which is the criterion for the 
lensing zone of central perturbations, varies depending on the 
mass ratio.  The dependency is such that a heavier planet has a 
wider lensing zone.

In Table~\ref{table:one}, we list lensing zones for several cases 
of planets.  We find that the lensing zone of central perturbations 
is equivalent to the classical lensing zone for a planet with a 
mass ratio $q\sim 3 \times 10^{-4}$.  For planets with mass ratios 
greater than this, the lensing zone of central perturbations is 
wider than the classical lensing zone.  For example, the lensing 
zone of a planet with $q=3 \times 10^{-3}$, which corresponds to 
a Jupiter-mass planet around a low-mass star with $M\sim 0.3\ 
M_\odot$, ranges from $0.25 \lesssim s \lesssim 3.9$. In physical 
units, this range corresponds to $0.5\ {\rm AU}\lesssim d \lesssim 
7.4\ {\rm AU}$ for a typical Galactic bulge event.  Considering 
the mass of the primary star, this lensing zone is equivalent to 
the range encompassing Mars to Uranus in our solar system when 
the dimension is scaled by the mass ratio of the lens to the sun.

The wider lensing zone of giant planets implies that microlensing 
method provides an important tool to detect planetary systems 
composed of multiple ice-giant planets.  This is because all 
planets located within the lensing zone will produce their own 
signatures at a common area of the central perturbation 
\citep{gaudi98, han05}.  In this sense, the recent discovery 
of a planetary system with a Jupiter/Saturn analog by the 
microlensing method \citep{gaudi08} is not a fluke and we 
predict that microlensing will discover more of such multiple 
planetary systems in the future if they are common.

\section{Conclusion}

While the convention of defining the range of planet separation 
detectable by the microlensing method is based on the location 
of the planet-induced caustic, the range for planets detectable 
through the channel of high-magnification events cannot be defined 
by the conventional criterion because the caustic is always 
located close to the primary lens regardless of the planet 
separation. Instead, the most important restriction to the 
detection of central perturbation is given by the finite-source 
effect. We investigated the magnification pattern in the central 
region of various planetary systems and found that perturbations 
with $\geq 5\%$ can be detected if the caustic is bigger than 
roughly one fourth of the caustic size.  Based on this finding, 
we derived an analytic expression for the range of planetary 
separations detectable through the central perturbations. From 
the comparison of classical lensing zone, we found that the 
new lensing zone is equivalent to the classical zone for a planet 
with a planet/star mass ratio $q\sim 3\times 10^{-4}$ and it 
becomes wider for heavier planets.  The wider lensing zone of 
the central perturbations for giant planets implies that the 
microlensing method provides an important tool to detect planetary 
systems composed of multiple ice-giant planets.

\acknowledgments 
This work was supported 
by the Astrophysical Research Center for the Structure and Evolution 
of the Cosmos (ARCSEC) of Korea Science and Engineering Foundation 
(KOSEF) through Science Research Program (SRC) program.


\begin{thebibliography}{99}
\frenchspacing

\bibitem[Abe et al.(2004)]{abe04}
Abe, F., et al.\ 2004, Science, 305, 1264

\bibitem[Beaulieu et al.(2006)]{beaulieu06}
Beaulieu, J.\ P., et al.\ 2006, Nature, 439, 437

\bibitem[Bennett \& Rhie(1996)]{bennett96}
Bennett, D.\ P., \& Rhie, S.\ H.\ 1996, \apj, 472, 660

\bibitem[Bennett et al.(2008)]{bennett08}
Bennett, D.\ P., et al.\ 2008, \apj, submitted

\bibitem[Bond et al.(2002a)]{bond02a}
Bond, I.\ A., et al.\ 2002, \mnras, 331, L19

\bibitem[Bond et al.(2002b)]{bond02b}
Bond, I.\ A., et al.\ 2002, \mnras, 333, 71

\bibitem[Bond et al.(2004)]{bond04}
Bond, I.\ A., et al.\ 2004, \apj, 606, L155

\bibitem[Chung et al.(2005)]{chung05}
Chung, S.-J., et al.\ 2005, \apj, 630, 535

\bibitem[Dominik(1999)]{dominik99}
Dominik, M.\ 1999, \aap, 349, 108

\bibitem[Dong et al.(2006)]{dong06}
Dong, S., et al.\ 2006, \apj, 642, 842

\bibitem[Dong et al.(2008)]{dong08}
Dong, S., et al.\ 2008, in preparation

\bibitem[Gaudi, Naber, \& Sackett(1998)]{gaudi98}
Gaudi, B.\ S., Naber, R.\ M., \& Sackett, P.\ D.\ 1998, \apj, 502, L33

\bibitem[Gaudi et al.(2008)]{gaudi08}
Gaudi, B.\ S., et al.\ 2008, Science, 319, 927

\bibitem[Gould \& Loeb(1992)]{gould92}
Gould, A., \& Loeb, A.\ 1992, \apj, 396,

\bibitem[Gould et al.(2006)]{gould06}
Gould, A., et al.\ 2006, \apj, 644, L37

\bibitem[Griest \& Safizadeh(1998)]{griest98}
Griest, K., \& Safizadeh, N.\ 1998, \apj, 500, 37

\bibitem[Han(2005)]{han05}
Han, C.\ 2005, \apj, 629, 1102

\bibitem[Han(2006)]{han06}
Han, C.\ 2006, \apj, 638, 1080

\bibitem[Kubas et al.(2008)]{kubas08}
Kubas, D., et al.\ 2008, \aap, 483, 317

\bibitem[Rattenbury et al.(2002)]{rattenbury02}
Rattenbury, N.\ J., Bond, I.\ A., Skuljan, J., \& Yock, P.\ C.\ M.\ 
2002, \mnras, 335, 159

\bibitem[Udalski(2003)]{udalski03}
Udalski, A.\ 2003, Acta Astron., 53, 291

\bibitem[Udalski et al.(2005)]{udalski05}
Udalski, A., et al.\ 2005, \apj, 628, L109

\bibitem[Wambsganss(1997)]{wambsganss97}
Wambsganss, J.\ 1997, \mnras, 284, 172

\end{thebibliography}
\end{document}